\documentclass[12 pt,aps,a4paper]{article}
\usepackage{enumerate}
\usepackage{amssymb}
\usepackage[latin1]{inputenc}
\usepackage{diagbox}
\usepackage{setspace}
\usepackage[english]{babel}
\usepackage{a4wide}
\usepackage{graphicx}
\usepackage{float}
\usepackage{amsthm}
\usepackage{graphics}
\usepackage{xcolor}
\usepackage{setspace}
\usepackage{subcaption}
\usepackage{mathtools}

\usepackage{afterpage}
\usepackage{indentfirst}
\usepackage{braket}
\usepackage{bm}
\usepackage{microtype}
\usepackage{wasysym}
\usepackage{array}
\usepackage{amsmath}

\begin{document}

\title{Thermodynamics of ancilla-assisted erasure of quantum information}

\date{\today}

\maketitle
\begin{center}

\author{Carlos Octavio A. Ribeiro Neto and Bert\'{u}lio de Lima Bernardo$^{*}$}

\end{center}

\begin{center}
Departamento de F\'{\i}sica, Universidade Federal da Para\'{\i}ba, 58051-900 Jo\~ao Pessoa, PB, Brazil
\end{center}

\begin{center}
*Corresponding author. E-mail: bertulio.fisica@gmail.com \\
Telephone Number: +55 83 99665 3997
\end{center}

\begin{abstract}
Landauer's principle sets a fundamental limit on the heat dissipated when one classical bit of information is erased from a memory, thereby establishing a direct link between information theory and thermodynamics. With the advent of quantum technologies, a natural question arises: how does Landauer's principle extend to the quantum regime? In this work, we study the thermodynamics of a quantum channel that erases an arbitrary state of a qubit memory in contact with a reservoir composed of a thermal qubit and a pure ancilla qubit. The channel is based only on CNOT gates, and the introduction of the ancilla makes it capable of operating beyond Landauer's limit, when the temperature of the thermal qubit is above a given limit temperature. However, we observe that, since the introduction of the ancilla does not correspond to a strict Landauer's scenario, our protocol does not represent a violation of Landauer's principle.  
\end{abstract}


                        
\maketitle


\section{\label{sec:level1}Introduction}

In 1961, Rolf Landauer demonstrated a fundamental limit on the amount of energy required to erase information from a classical system \cite{landauer}. This limit, which is now called Landauer's principle, states that the heat $Q_E$ dissipated to the environment in an erasure process obeys the lower bound \cite{maru}: 
\begin{equation}
\label{landauer}
Q_E \geq k_B T \Delta S,
\end{equation}
where $\Delta S = S_i - S_f$ is the entropy decrease of the system, with $S_i$ and $S_f$ being the initial and final entropies, respectively, $T$ is the temperature of the environment with which the system exchanges heat, and $k_B$ is the Boltzmann constant. This relation establishes that the lowest possible energy cost to erase a classical bit is $k_B T \, \mathrm{ln}  \, 2$. From both the fundamental and practical viewpoints, Landauer's bound is of central importance as it represents a link between information theory and thermodynamics, and serves as a reference to assess the performance of information processing systems. Experimentally, the principle was first verified using a single colloidal particle trapped in an optical tweezer-based double-well potential \cite{berut}. Later, it was confirmed using nanomagnetic memory bits \cite{hong}.  

The past decade has witnessed a rapid growth of interest in quantum computing technologies and energy-based quantum devices
\cite{preskill,kjaer, klat,bouton,kim,auff,koch}. Then, with the increasing need to design and better describe the operation of such architectures, we were led to a scenario that required the development of the theory of quantum thermodynamics, which investigates thermodynamic processes at the quantum level \cite{goold,deffner,binder}. In this perspective, it is natural to ask whether and how the Landauer principle applies to quantum dynamics. This issue has been extensively studied from both the theoretical \cite{reeb, timp,vu,taranto,bis} and experimental \cite {silva,yan} points of view. In Ref. \cite{allah}, the authors pointed out the possibility of violation when a quantum system gets entangled with the environment, as long as a direct identification between the Landauer principle and the Clausius inequality is considered. However, the former has been shown to be derived independently of the latter \cite{shiz,piech}. More recently, it was proposed a relation between the violation of Landauer's principle and the information backflow that occurs in a system-environment non-Markovian dynamics  \cite{pezzuto,man,hu}.        

In the classical framework \cite{landauer}, the erasure process consists of an engineered manipulation of a {\it memory system} in contact with a {\it thermal reservoir}, in which the memory is reliably reset to a {\it definite} state at the end. With this motivation, Reeb and Wolf proposed that an equivalent quantum-mechanical scenario for Landauer's erasure principle should be based on four assumptions \cite{reeb}: i) the process involves a {\it memory} $M$ and a {\it reservoir} $R$, individually described by Hilbert spaces; ii) the reservoir $R$ is initially in a thermal state; iii) the memory $M$ and the reservoir $R$ are initially uncorrelated, $\hat{\rho}^{(i)} = \hat{\rho}^{(i)}_{M} \otimes \hat{\rho}^{(i)}_{R}$; iv) the erasure process occurs via an unitary operation involving the memory and the reservoir, $\hat{\rho}^{(f)} = \hat{\mathcal{U}} (\hat{\rho}^{(i)}_{M} \otimes \hat{\rho}^{(i)}_{R})\hat{\mathcal{U}}^{\dagger} = \hat{\mathcal{U}}\hat{\rho}^{(i)}\hat{\mathcal{U}}^{\dagger}$. Here, we consider the memory as the main system undergoing erasure.     

In this work, we propose a quantum channel that efficiently erases any state of a qubit memory by putting it in contact with a thermal qubit at finite temperature, as prescribed by the above assumptions. Nevertheless, we relax assumption ii) by including an {\it ancilla} qubit $A$ prepared in a pure state, such that our reservoir is considered as the thermal qubit plus the ancilla. This ancillary qubit can be either an independent system or a reservoir degree of freedom other than the energy. Both cases are operationally equivalent, but we shall discuss the second one with more detail using an atom-atom interaction scenario. Using our quantum channel, we are able to show that it operates in a regime beyond the Landauer limit when the temperature of the thermal qubit is higher than a given limit temperature. However, since we relax one of Landauer's scenario assumptions, we cannot say that it represents a violation of Landauer's principle. We present the erasure protocol in Sec.~\ref{erasure}. In Sec.~\ref{thermodynamics}, we discuss the thermodynamics of the protocol from both the entropic and the energetic viewpoints. Sec.~\ref{landlimit} is dedicated to the study of how our quantum channel fits within Landauer's scenario. An optical simulation of our erasure protocol is depicted in Sec.~\ref{opticalsimulation}. Finally, our concluding remarks are presented in Sec.~\ref{conclusions}.

\section{Erasure Protocol} \label{erasure}

\begin{figure}[h!]
    \centering
\includegraphics[scale=0.27]{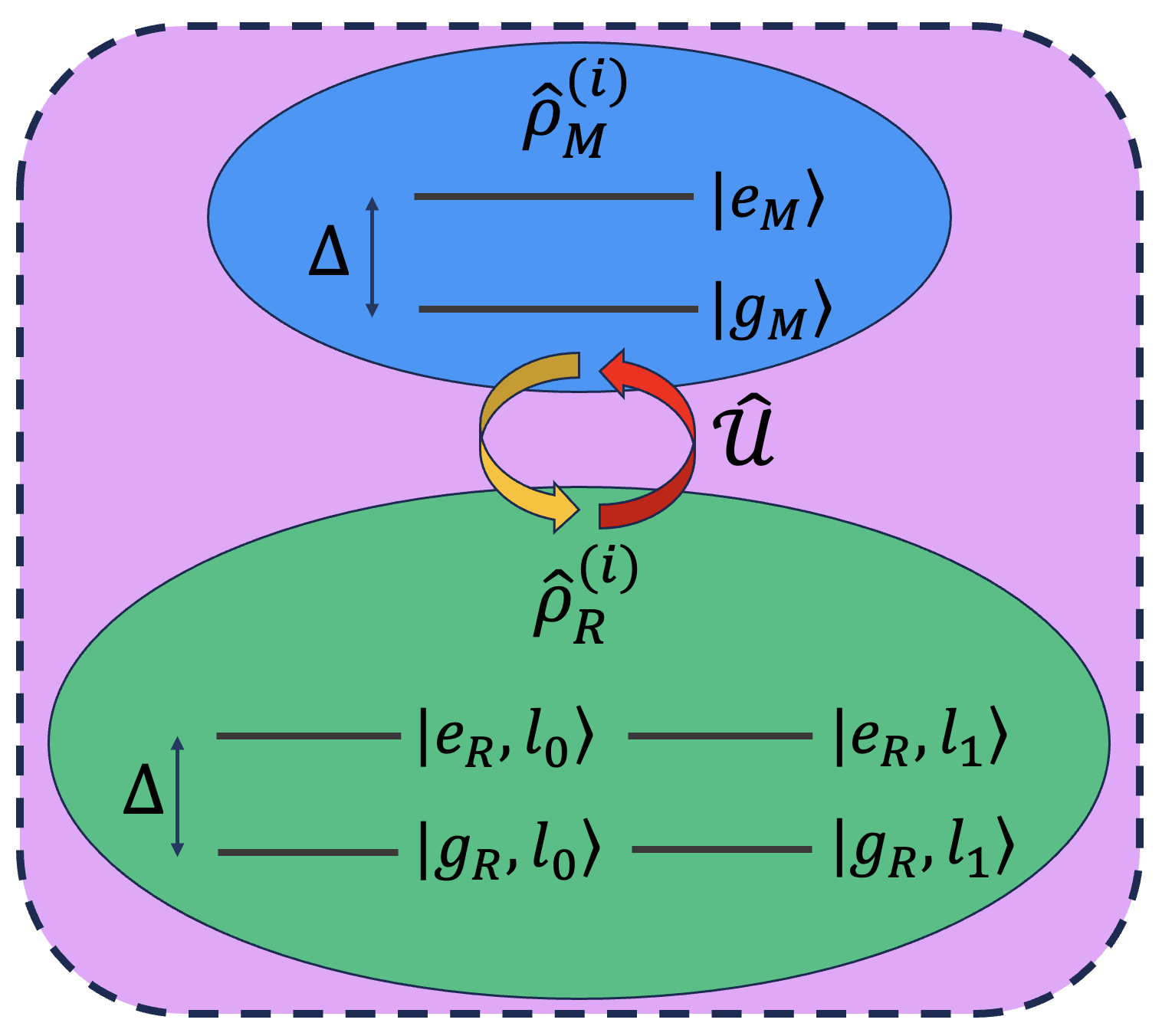}
\caption{Representation of the composite system in the initial 
 state $\hat{\rho}^{(i)}$, which consists of a qubit in the state $\hat{\rho}^{(i)}_M$ as the memory and a reservoir with two two-fold degenerate states in the state $\hat{\rho}^{(i)}_R$. Both have the same energy-level spacing $\Delta$, and the unitary transformation $\hat{\mathcal{U}}$ determines their joint evolution.}
    \label{fig1}
\end{figure}

\begin{figure}[h!]
    \centering
    \includegraphics[scale=0.32]{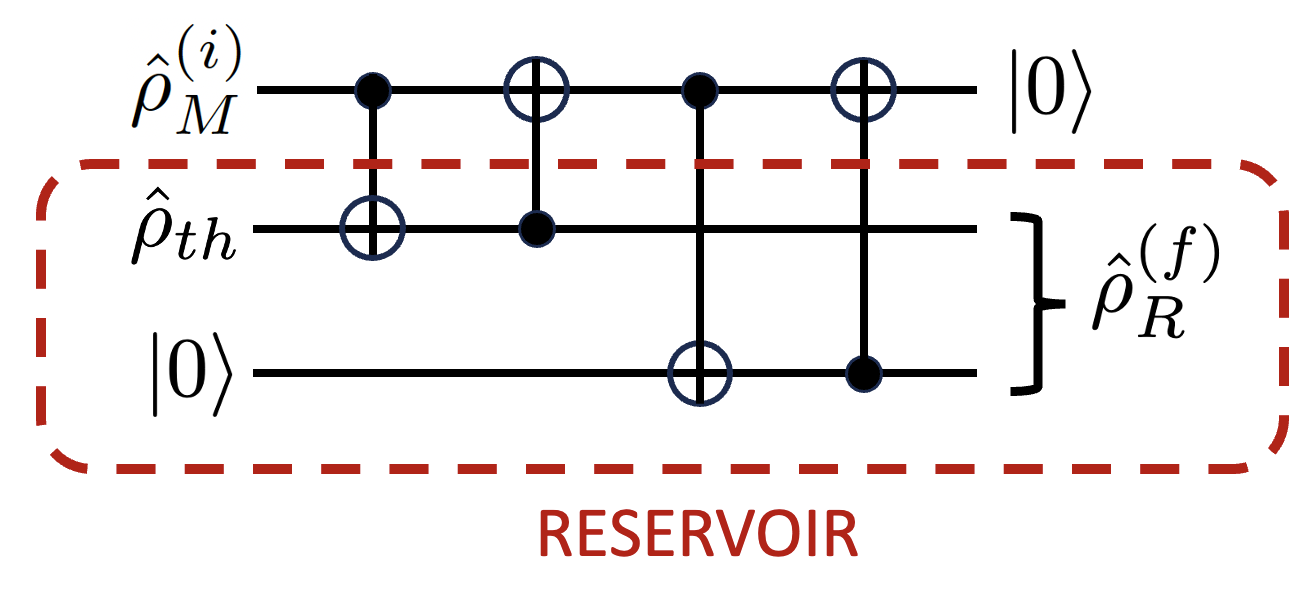}
    \caption{Quantum circuit diagram for the erasure protocol, which is made up of four CNOT gates. The top line is the memory, the central line is the energy of the reservoir and the bottom line the ancilla. To better approach the Landauer scenario, we assume that the memory starts in an arbitrary qubit state $\hat{\rho}^{(i)}_{M}$, and the reservoir energy in the thermal state $\hat{\rho}_{th}$. The ancillary qubit is required to be prepared in the state $\ket{0}$. At the end, the memory always gets erased, i.e., it is transformed into the state $\ket{0}$, whereas the two other qubits, which comprise the reservoir, evolve into the final state $\hat{\rho}^{(f)}_{R}$.}
    \label{fig2}
\end{figure}

In the quantum mechanical domain, an effective erasure process is expected to map any arbitrary state of a memory system into a well-defined pure state. This transformation is necessarily accompanied by a flux of entropy between the memory that stores information and a reservoir, which is responsible for receiving the information that results from the erasure. We bring attention to the fact that we define the {\it environment} and the {\it reservoir} as different physical entities. Here, the environment comprises the entire universe, except memory, whereas the reservoir is the portion of the environment that directly participates in the erasure operation. To introduce our idea, we first consider the atom-atom interaction scenario schematically represented in Fig.~\ref{fig1}, and subsequently discuss the corresponding quantum circuit, shown in Fig.~\ref{fig2}. The memory is considered as a qubit system, whose state can be spanned by the energy basis $\{ \ket{g_{M}},\ket{e_{M}}\}$, with a free Hamiltonian given by
\begin{equation}\label{eq:1}
    \hat{H}_{M} = E\ket{g_{M}}\bra{g_{M}} + (E + \Delta)\ket{e_{M}}\bra{e_{M}}.
\end{equation}
The reservoir is a quantum system with two two-fold degenerate energy states denoted by $\{\ket{g_{R}, l_{0}}, \ket{g_{R}, l_{1}}\}$ and $\{\ket{e_{R}, l_{0}}, \ket{e_{R}, l_{1}}\}$. They represent two pairs of states, where each pair shares the same energy state ($\ket{g_R}$ or $\ket{e_R}$), but different angular momentum (AM) states ($\ket{l_0}$ or $\ket{l_1}$). The Hamiltonian of the reservoir is assumed to be
\begin{multline}\label{eq:2}
    \hat{H}_{R} = \epsilon(\ket{g_{R}, l_{0}}\bra{g_{R}, l_{0}} + \ket{g_{R}, l_{1}}\bra{g_{R}, l_{1}}) + 
    (\epsilon + \Delta)(\ket{e_{R}, l_{0}}\bra{e_{R}, l_{0}} + \ket{e_{R}, l_{1}}\bra{e_{R}, l_{1}}),
\end{multline}
where the energy difference between the two pairs of levels $\Delta$ is the same as that of the two states of the memory [see Eq.~(\ref{eq:1})]. Above, $\ket{g}$ and $\ket{e}$ are used to denote the ground and the excited energy states of the memory and the reservoir.

Our erasure protocol is described by the following set of transformations of the memory and the reservoir
\begin{equation}\label{first}
        \ket{g_{M};g_{R},l_{0}} \longrightarrow \ket{g_{M};g_{R},l_{0}},
    \end{equation}
    \begin{equation}\label{second}
        \ket{g_{M};e_{R},l_{0}} \longrightarrow \ket{g_{M};e_{R},l_{1}},
    \end{equation}
      \begin{equation}\label{third}
        \ket{e_{M};g_{R},l_{0}} \longrightarrow \ket{g_{M};e_{R},l_{0}},
    \end{equation}
    \begin{equation}\label{fourth}
        \ket{e_{M};e_{R},l_{0}} \longrightarrow \ket{g_{M},g_{R},l_{1}}.
    \end{equation}
All processes above lead the memory to the state $\ket{g_{M}}$, along with some transformation in the reservoir state. 

In the present context, Eqs.~(\ref{first}) to~(\ref{fourth}) can be physically interpreted as follows: i) Eq.~(\ref{first}) says that when the memory and the reservoir are both in the ground state, and the latter has AM $l_{0}$, no change takes place. ii) Eq.~(\ref{second}) says that with the memory in the ground state and the reservoir in the excited state with AM $l_{0}$, their interaction is such that the only change is that the AM turns into $l_{1}$. This change in the AM state could result from a memory-reservoir collision, being compensated by a change in the linear momentum of the composite system. iii) Eq.~(\ref{third}) tells us that if the memory is in the excited state, with the reservoir in the ground state and AM $l_0$, the memory emits a photon which is absorbed by the reservoir, so that the former ends up in the ground state and the latter in the excited state. The conservation of the AM as $l_0$, despite the photon absorption, can also be a product of a collision that results in a change of the composite system's linear momentum. iv) Eq.~(\ref{fourth}) states that when the memory and the reservoir are both in the excited state, with the latter with AM $l_{0}$, the memory emits a photon which induces a stimulated emission in the reservoir. As a consequence, both end up in the ground state, the AM turns into $l_{1}$, and two photons are created in the process.

The above set of transformations can be described as part of the following unitary operation
\begin{equation}\label{unitary}
    \mathcal{\hat{U}} = 
    \begin{bmatrix}
    1 & 0 & 0 & 0 & 0 & 0 & 0 & 0 \\
    0 & 0 & 0 & 0 & 0 & 0 & 1 & 0 \\
    0 & 0 & 0 & 0 & 1 & 0 & 0 & 0 \\
    0 & 0 & 1 & 0 & 0 & 0 & 0 & 0 \\
    0 & 0 & 0 & 0 & 0 & 0 & 0 & 1 \\
    0 & 1 & 0 & 0 & 0 & 0 & 0 & 0 \\
    0 & 0 & 0 & 1 & 0 & 0 & 0 & 0 \\
    0 & 0 & 0 & 0 & 0 & 1 & 0 & 0 \\
    \end{bmatrix},
\end{equation}
which is written in the representation basis 
\begin{align} 
\{\ket{g_{M};g_{R},l_{0}},
\ket{g_{M};g_{R},l_{1}}, \ket{g_{M};e_{R},l_{0}}, \ket{g_{M};e_{R},l_{1}}, \nonumber
\\
\ket{e_{M};g_{R},l_{0}},
\ket{e_{M};g_{R},l_{1}}, \ket{e_{M};e_{R},l_{0}}, \ket{e_{M};e_{R},l_{1}}\}. 
\end{align}
The unitary $\mathcal{\hat{U}}$ describes the interaction between the memory and the reservoir in our erasure protocol,
\begin{equation}\label{transformation}
    \hat{\rho}^{(i)} \longrightarrow \hat{\rho}^{(f)} = \hat{\mathcal{U}}\hat{\rho}^{(i)}\hat{\mathcal{U}}^{\dagger},
\end{equation}
where $\hat{\rho}^{(i)}$ and $\hat{\rho}^{(f)}$ represent the initial and final states of the composite system, respectively.

Here, we consider the memory and the  reservoir initially in a product state,
\begin{align}
\label{icstate}
    \hat{\rho}^{(i)} &= \hat{\rho}^{(i)}_{M} \otimes \hat{\rho}^{(i)}_{R}.
\end{align}
In this case, $\hat{\rho}^{(i)}_{M}$ is the arbitrary initial state of the memory to be erased, and $\hat{\rho}^{(i)}_{R}$ is the initial state of the reservoir. Since the memory is a qubit system, we can write $\hat{\rho}_{M}^{(i)}$ as the general state \cite{nielsen} 
\begin{equation}
\label{imstate}
\hat{\rho}^{(i)}_{M} = \frac{\hat{\mathbb{I}}_M + \Vec{r}\cdot\Vec{\mathbf{\sigma}}}{2} = \frac{1}{2} (\hat{\mathbb{I}}_M + r_x \hat{\sigma}_{x} + r_y \hat{\sigma}_{y} + r_z \hat{\sigma}_{z}),
\end{equation}
where $\hat{\mathbb{I}}_M$ is the identity operator in the memory Hilbert space, $\Vec{r} = (r_{x},r_{y},r_{z})$ is the Bloch vector, and $\Vec{\sigma} =(\hat{\sigma}_{x},\hat{\sigma}_{y},\hat{\sigma}_{z})$, with $\hat{\sigma}_{x}$, $\hat{\sigma}_{y}$, and $\hat{\sigma}_{z}$ being the Pauli operators. The components of the Bloch vector are given by $r_\mu = \mathrm{Tr}(\hat{\rho}^{(i)}_{S} \sigma_\mu)$, with $\mu = x,y$ and $z$, and Tr stands for the trace operation.

In order to better approach a quantum Landauer scenario \cite{reeb}, the  state of the reservoir is assumed to be originally a thermal state with inverse temperature $\beta = 1/k_B T$,
\begin{eqnarray}
    \label{thermal}
    \hat{\Gamma}_{th} = \frac{1}{2} \Bigl[ p_{g} \ket{g_{R}, l_{0}}\bra{g_{R}, l_{0}} + p_{g} \ket{g_{R}, l_{1}}\bra{g_{R}, l_{1}} \nonumber 
    \\
    + p_{e}\ket{e_{R}, l_{0}}\bra{e_{R}, l_{0}}  + p_{e}\ket{e_{R}, l_{1}}\bra{e_{R}, l_{1}}\Bigl],
\end{eqnarray}
where $p_g$ and $p_e$ are the occupation probabilities of the ground and excited levels, which are described by the Boltzmann distribution, $p_{e} = e^{-\beta \Delta} p_{g}$ and $p_{g} + p_{e} = 1$. The factor 1/2 is due to the degeneracy of levels. 

Before the interaction with the memory, we require that this thermal state is preselected with AM $l_0$. This preselection of the reservoir into the  state $\ket{l_{0}}$ ensures that only the memory-reservoir interactions described in Eqs.~(\ref{first}) to~(\ref{fourth}) will take place. This operation allows us to write the resulting reservoir state as
\begin{align}
\label{irstate}
\hat{\rho}^{(i)}_{R} &= p_{g} \ket{g_{R}, l_{0}}\bra{g_{R}, l_{0}} + p_{e}\ket{e_{R}, l_{0}}\bra{e_{R}, l_{0}} \nonumber \\
&=\hat{\rho}_{th} \otimes \ket{l_0}\bra{l_0},
\end{align}
where $\hat{\rho}_{th} = p_{g} \ket{g_{R}}\bra{g_{R}} + p_{e}\ket{e_{R}}\bra{e_{R}}$ is a thermal state of the energy degree of freedom of the reservoir, and $\ket{l_0}$ is the preselected AM state.
As we shall see, the AM degree of freedom serves as a pure ancilla qubit in the erasure process (cf. Fig.~\ref{fig2}).  Here, $\hat{\rho}^{(i)}_{R}$ is considered as our initial state of the reservoir. We can observe that, in the atomic context, the AM preselection could be realized with a Stern-Gerlach apparatus \cite{sakurai}.

If we calculate the transformation given in Eq.~(\ref{transformation}), by using Eqs.~(\ref{unitary}),~(\ref{icstate}),~(\ref{imstate}) and~(\ref{irstate}), we obtain the final composite system state as given by 
\begin{equation}
\label{fcstate}
    \hat{\rho}^{(f)} = \ket{g_{M}}\bra{g_{M}} \otimes \hat{\rho}^{(f)}_{R},
\end{equation}
where
\begin{align}
\label{frstate}
\hat{\rho}^{(f)}_{R} &= \frac{1}{2} \Bigl[ (1 + r_{z})(p_{g}\ket{g_{R},l_{0}}\bra{g_{R},l_{0}} + p_{e}\ket{e_{R},l_{1}}\bra{e_{R},l_{1}}) \nonumber
\\
    &+ (r_{x} - ir_{y})(p_{g}\ket{g_{R},l_{0}}\bra{e_{R},l_{0}} +
p_{e}\ket{e_{R},l_{1}}\bra{g_{R},l_{1}}) \nonumber
\\
    &+ (r_{x} + ir_{y})(p_{g}\ket{e_{R},l_{0}}\bra{g_{R},l_{0}} +
p_{e}\ket{g_{R},l_{1}}\bra{e_{R},l_{1}}) \nonumber
\\
    &+ (1 - r_{z})(p_{g}\ket{e_{R},l_{0}}\bra{e_{R},l_{0}} 
 +p_{e}\ket{g_{R},l_{1}}\bra{g_{R},l_{1}}) \Bigr]
\end{align}
is the final state of the reservoir. As can be seen, the final state of the composite system in Eq.~(\ref{fcstate}) is uncorrelated, and the final state of the memory is $\ket{g_M}$, independently of the initial state $\hat{\rho}^{(i)}_{M}$. That is to say that the process erases the quantum state of the memory with unit probability, with the state of the reservoir being transformed into that given in Eq.~(\ref{frstate}). It can be verified that the state of Eq.~(\ref{frstate}) displays no quantum correlations. However, the terms involving $r_{x} \pm i r_{y}$ certifies the presence of coherence in the reservoir energy whenever $r_{x}$ or $r_{y}$ are nonzero. This means that if the initial state of the memory $\hat{\rho}^{(i)}_{M}$ has some coherence, this coherence is transferred to the energy degree of freedom of the reservoir after the erasure protocol.

To further clarify the details of our erasure protocol, and make possible the implementation with different quantum platforms, in Fig.~\ref{fig2} we depict the corresponding quantum circuit. The top line is the memory qubit, and the other two lines the reservoir. Specifically, the intermediate line represents the reservoir energy and the bottom line the reservoir AM (ancilla qubit). The memory starts in the arbitrary qubit state $\hat{\rho}^{(i)}_{M}$ and is transformed into the {\it erased} state $\ket{g_M}$ ($\ket{0}$). The reservoir energy starts in the thermal state $\hat{\rho}_{th}$, whereas the AM is prepared in the pure state $\ket{l_0}$ ($\ket{0}$). The reservoir, which comprises the energy and the AM degrees of freedom, ends up in the state $\hat{\rho}^{(f)}_{R}$. As can be seen, the complete erasure channel can be built from just four CNOT gates.

From the structure of the quantum circuit shown in Fig.~\ref{fig2}, we observe that the ancilla, above described as the reservoir AM, could otherwise be an independent qubit system. This means that the erasure protocol can also be realized by taking a thermal qubit and a pure ancilla qubit independently, which together play the role of the reservoir. One point to be noted in the circuit of Fig.~\ref{fig2} is that, despite the transfer of the initial state $\ket{0}$ from the ancilla to the memory, it is conceptually different from the SWAP process \cite{reeb,silva}. This is because the latter case restricts the memory to interact with a zero-temperature reservoir in order to be reset into the state $\ket{0}$ at the end. 

Before studying the thermodynamics of our erasure protocol in the next section, we should comment on the unitarity property of the general process described by $\mathcal{\hat{U}}$, which is one of the assumptions of the quantum Landauer scenario presented in the Introduction. We highlight that a proper erasure process must be irreversible, meaning that no information about the initial state of the memory can be recovered from the reservoir once the process is complete. This raises a natural question: why is the erasure process modeled by a unitary operation, given that unitarity implies reversibility via the application of $\mathcal{\hat{U}}^{\dagger}$? In the unitary operation, no {\it environment} constituent, other than the {\it thermal qubit} and the {\it ancilla}, interacts with the {\it memory} \cite{reeb}. However, in reality, the erasure process is not complete with the unitary operation.

To complete the erasure, after the unitary, the final state of the reservoir $\hat{\rho}^{(f)}_{R}$ has to be erased, since it still holds records of the initial state of the memory, $\hat{\rho}^{(i)}_{M}$. In practice, the erasure of $\hat{\rho}^{(f)}_{R}$ takes place naturally upon subsequent interactions of the two qubits of the reservoir with the other constituents of the environment, in such a way that the information about $\hat{\rho}^{(i)}_{M}$ spreads out and becomes irretrievable. As a matter of fact, such subsequent interactions play no role in the investigation of the Landauer bound because they do not involve the memory. Put differently, to calculate the heat dissipated $Q_E$ and the entropy decrease $\Delta S$, which are the quantities involved in Ineq.~(\ref{landauer}), we only need to consider the interactions that involve the memory, i.e., those described by the unitary operation $\mathcal{\hat{U}}$.

\section{Thermodynamics of the protocol} \label{thermodynamics}

The study of entropy dynamics and energy exchange plays an important role in the description of quantum thermodynamic processes \cite{vinj,bert,strasberg}. In this section, we will investigate our erasure operation from both perspectives, and in the next section proceed to discuss how it fits into Landauer's principle.

\subsection{Entropy analysis}

Here, we use the von Neumann entropy to quantify the amount of information present in a quantum system. For a system in a quantum state $\hat{\rho}$, this is given by $S(\hat{\rho}) = - \mathrm{Tr}(\hat{\rho}\mathrm{ln}\hat{\rho})$, where we use the natural logarithm, such that it is measured in {\it nats} ($\approx 1.44$ bits). We observed in Eqs.~(\ref{icstate}) and~(\ref{fcstate}) that the composite system is in a product state at the beginning and at the end of the erasure operation. This implies that in both cases the entropy of the composite system is given simply by the sum of the individual entropies of the memory and the reservoir \cite{nielsen}. Furthermore, we have that $\hat{\rho}_{M}^{(f)}$ is always the pure state $\ket{g_M}$, which means that $S(\hat{\rho}_{M}^{(f)}) = 0$. We also have that $ S(\hat{\rho}^{(f)}) = S(\hat{\rho}^{(i)})$ because the composite system undergoes an unitary transformation. Therefore, we have that the entropy initially contained in the memory with the state $\hat{\rho}_{M}^{(i)}$ is completely transferred to the reservoir after the erasure operation, which ends up in the state $\hat{\rho}^{(f)}_{R}$.

With the above considerations, we can define the entropy decrease of the memory as
\begin{equation}\label{edecrease}
    \Delta S = S(\hat{\rho}^{(i)}_{M}) - S(\hat{\rho}^{(f)}_{M})= S(\hat{\rho}^{(i)}_{M}).
\end{equation}
By substitution of Eq.~(\ref{imstate}) into Eq.~(\ref{edecrease}), after some calculations, we find that
\begin{equation}\label{edecrease2}
     \Delta S = \mathrm{ln}2 - r\mathrm{ln}(1+r) - \frac{1}{2}(1-r)\mathrm{ln}(1-r^{2}),
\end{equation}
where $r = \sqrt{r^2_x + r^2_y + r^2_z}$, with $0 \leq r \leq 1$, is the length of the Bloch vector. It can be verified that $\Delta S$ is a concave function of $r$, with $\Delta S = 0$ when $r=1$ ($\hat{\rho}^{(i)}_{M}$ is a pure state), and $\Delta S = \mathrm{ln} 2$ when $r=0$ ($\hat{\rho}^{(i)}_{M}$ is a completely mixed state).

Let us now comment on the physical importance of the ancillary AM degree of freedom of the reservoir, and the absence of the $\ket{l_1}$ state in $\hat{\rho}^{(i)}_{R}$, for the realization of the erasure operation. Since the initial entropy of the memory, which is at most $\mathrm{ln} 2$, is totally transferred to the reservoir, the absence of $\ket{l_1}$ in $\hat{\rho}^{(i)}_{R}$ guarantees that the reservoir will be capable of storing it. In other words, the occupation of $\ket{l_1}$ is initially suppressed in order to allow the ancilla to receive the information that comes from the memory.

\subsection{Energy analysis}

Since entropy is transferred from the memory to the reservoir in the erasure protocol, we also expect that heat is exchanged between them. Actually, all change in their internal energy is due to heat transfer because the energy levels are kept fixed in the process, which means that no work is done on or by them \cite{bert,quan}. Here, we consider heat as the energy transfer accompanied by both entropy and coherence change. Some works classify the last type as {\it coherent energy} (e.g., Refs.~\cite{bert,bert2,khan,junior}). Let us now study the energetic influence of the unitary operation $\hat{\mathcal{U}}$ on the memory and the reservoir. For the composite system, we define the internal energy before and after the operation as
\begin{equation}
\label{icsie}
U^{(i)} = \mathrm{Tr} [\hat{\rho}^{(i)} \hat{H}]
\end{equation}
and 
\begin{equation}
\label{fcsie}
U^{(f)} = \mathrm{Tr} [\hat{\rho}^{(f)} \hat{H}],
\end{equation}
respectively, with $\hat{H} = \hat{H}_{M} \otimes \hat{\mathbb{I}}_{R} + \hat{\mathbb{I}}_{M} \otimes \hat{H}_{R}$ being the Hamiltonian of the composite system before and after the process, i.e., when there is no interaction between the memory and the reservoir. $\hat{\mathbb{I}}_{R}$ is the identity operator in the reservoir Hilbert space. By using the result of Eq.~(\ref{transformation}) into Eq.~(\ref{fcsie}) we can write
\begin{equation}
\label{fcsie2}
U^{(f)} = \mathrm{Tr} [\hat{\mathcal{U}}\hat{\rho}^{(i)}\hat{\mathcal{U}}^{\dagger} \hat{H}] = \mathrm{Tr} [\hat{\rho}^{(i)}\hat{\mathcal{U}}^{\dagger} \hat{H} \hat{\mathcal{U}}],
\end{equation}
where we used the cyclic invariance of the trace in the second equality \cite{sakurai,nielsen}. 

For the transformation caused by $\hat{\mathcal{U}}$ on the composite system to be considered energy-preserving, it is necessary that $U^{(f)} = U^{(i)}$, which from Eqs.~(\ref{icsie}) and~(\ref{fcsie2}) means that
$\hat{\mathcal{U}}^{\dagger} \hat{H} \hat{\mathcal{U}} = \hat{H}$, or simply that $[\hat{\mathcal{U}}, \hat{H}] = 0$. However, a direct calculation of this commutator shows that $[\hat{\mathcal{U}}, \hat{H}] \neq 0$, i.e., our erasure operation does not conserve the energy of the composite system. The physical reason for this is the fact that the transformation given in Eq.~(\ref{fourth}) does not conserve the energy, since the pair of photons created characterizes a dissipative process that occurs in the composite system. In contrast, the relations given in Eqs.~(\ref{first}) to~(\ref{third}) are all energy-preserving. 
 
Having analyzed the energy of the composite system, we now turn our attention to the memory and the reservoir separately. The heat exchanged by the memory in the process reads
\begin{align}
\label{hem}
   Q_{M} &= \Delta U_{M} = \mathrm{Tr}[(\hat{\rho}^{(f)}_{M}  - \hat{\rho}^{(i)}_{M}) \hat{H}_{M}] \nonumber \\
   &= -\frac{\Delta}{2} (1 - r_{z}).
\end{align}
For the reservoir, the heat exchanged can be shown to be
\begin{align}
\label{her}
    Q_{R} & = \Delta U_{R} = \mathrm{Tr}[(\hat{\rho}^{(f)}_{R} - \hat{\rho}^{(i)}_{R}) \hat{H}_{R}]\nonumber\\
    &= \frac{\Delta}{2} (1 - r_{z}) (p_g - p_e).
\end{align}
Above, $U_M$ and $U_R$ stand for the internal energies of the memory and the reservoir, respectively. Since  $-1 \leq r_{z} \leq 1$, we observe that $Q_{M} \leq 0$, which means that the memory {\it releases} heat. In turn, the fact that $p_g \geq p_e$ for thermal states (and hence for $\hat{\rho}^{(i)}_{R}$) implies that $ Q_{R} \geq 0$, which says that the reservoir {\it absorbs} heat. In both cases, the heat exchanged depends only on the component $r_z$ of the Bloch vector, which is the one associated with the internal energy of the memory.

The above results reveal interesting aspects of our erasure operation. For example, we observe that the heat released by the memory $Q_{M}$ is independent of the thermal qubit temperature. This can be explained from Eqs.~(\ref{first}) to~(\ref{fourth}). We observe that when the thermal qubit is in the ground state, Eq.~(\ref{first}) tells us that the memory has the energy unchanged if it is in the ground state, but Eq.~(\ref{third}) says that the energy decreases by an amount $\Delta$ if the memory starts in the excited state. In parallel, the same analysis holds true with respect to Eqs.~(\ref{second}) and~(\ref{fourth}) for the case in which the thermal qubit is initially in the excited state. Therefore, it is expected that the energy released by the memory is independent of the initial energy state of the thermal qubit, which explains the fact that $Q_{M}$ is temperature-independent. 

Other interesting points to be observed are that the heat absorbed by the reservoir $Q_{R}$ is temperature-dependent and that the reservoir does not absorb all the heat released by the memory, $Q_{R} \neq - Q_{M}$. 
The only exception to the latter statement is when $T=0$ ($p_g = 1$ and $p_g = 0$), which is the unique energy-preserving configuration of the composite system. That is to say that the general $Q_{R} \neq - Q_{M}$ behavior signifies that the composite system dissipates energy. Still, one can easily verify that the discrepancy between $Q_{R}$ and $- Q_{M}$ becomes more prominent for higher values of $T$ ($p_g \rightarrow 1/2^{-}$ and $p_e \rightarrow 1/2^{+}$). This is because in this regime the excited state of the thermal qubit becomes increasingly more populated, a fact that strengthens the influence of the dynamics described by Eq.~(\ref{fourth}). We remind that this is the dynamics responsible for making $\hat{\mathcal{U}}$ a non-energy-preserving operation. 

\section{Beyond Landauer's limit} \label{landlimit}

The study of the entropic and energetic properties of the erasure operation developed in the previous section puts us in a position to verify whether Landauer's bound is fulfilled. To do so, we use the temperature $T$ of the thermal qubit as a reference. The amount of information $\Delta S$ erased in the process was calculated in Eq.~(\ref{edecrease2}), and the heat dissipated by the memory $Q_M$ in Eq.~(\ref{hem}). One important point to observe here is that, given the initial state of the memory $\hat{\rho}^{(i)}_{M}$, we have that $\Delta S$ is a constant, and that $Q_M$ is temperature-independent. From the viewpoint of the memory, Ineq.~(\ref{landauer}) can be rewritten as
\begin{equation}
\label{landauermemo}
Q_M \leq - k_B T \Delta S,    
\end{equation}
where we used that $Q_M = - Q_E$. We recall that the {\it reservoir} (i.e., thermal qubit plus the ancilla) is the portion of the {\it environment} which directly participates in the erasure operation, as defined in Sec.~\ref{erasure}. In our protocol, we observe that $Q_E \geq Q_R$ because the heat absorbed by the reservoir does not encompass the energy of the pair of photons which may be created if the process described by Eq.~(\ref{fourth}) takes place.

\begin{figure}[h!]
    \centering
    \includegraphics[scale=0.25]{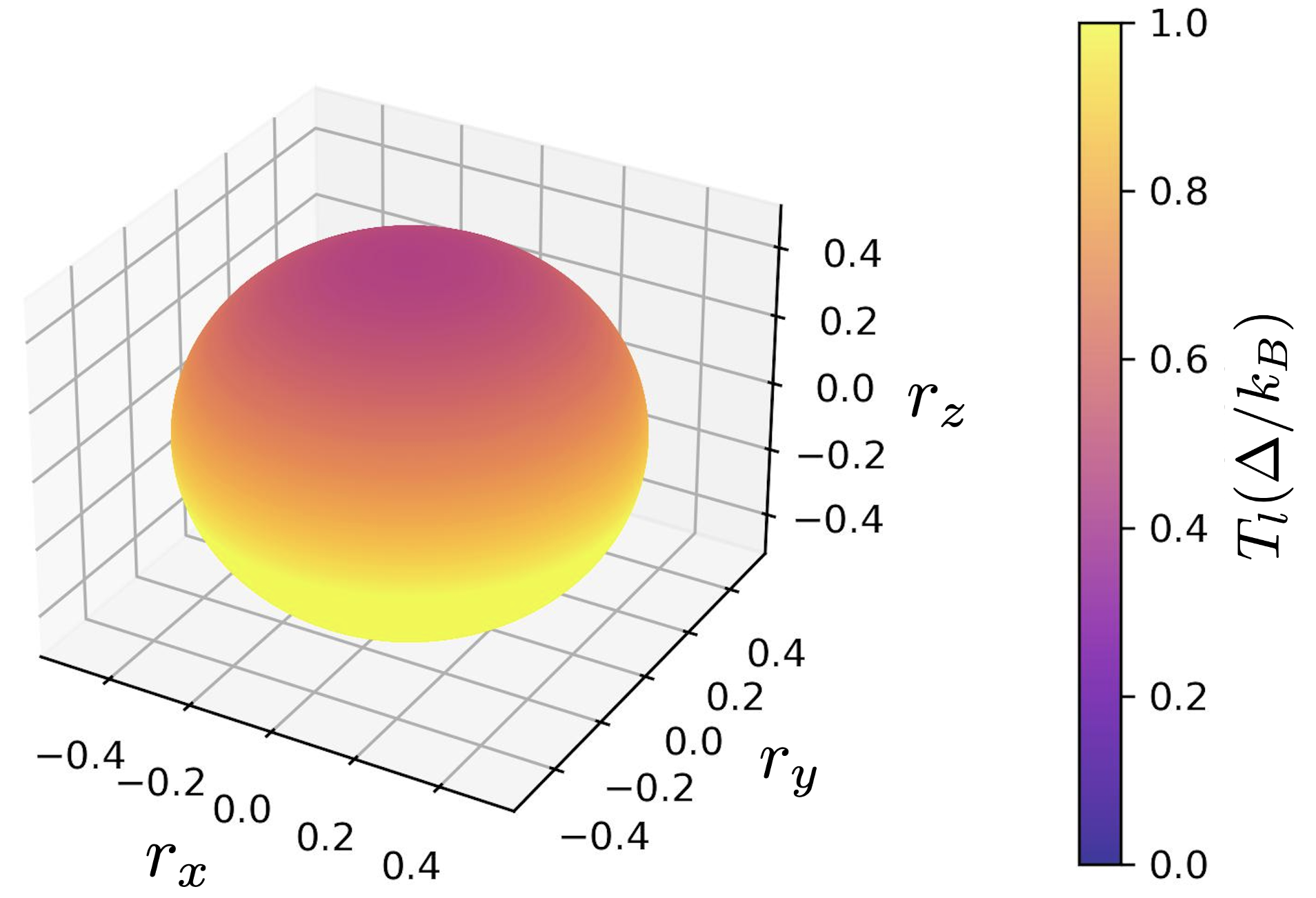}
    \caption{Color map of the limit temperature $T_l$, in units of $\Delta / k_B$, represented in the Bloch sphere for initial states of the memory $\hat{\rho}^{(i)}_{M}$ with $r=0.5$. We observe that lower-energy states ($r_z > 0$) correspond to lower values of $T_l$. This is because $-Q_M$ decreases linearly with $r_z$ in our protocol, and $\Delta S$ is the same positive constant for all states shown.}
    \label{temperature}
\end{figure}

By substitution of the results of Eqs.~(\ref{edecrease2}) and~(\ref{hem}) into the Landauer bound as shown in Ineq.~(\ref{landauermemo}), we observe that it is not fulfilled for temperatures {\it higher} than a limit temperature $T_l = - Q_M / (k_B \Delta S)$, which is given by    
\begin{equation}
\label{limit}
T_l = \frac{\Delta (1 - r_{z})}{k_B \left[\mathrm{ln} \, 4 - r\mathrm{ln}(1+r)^2 - (1-r)\mathrm{ln}(1-r^{2}) \right]}.
\end{equation}
The fact that Landauer's bound is not satisfied for $T > T_l$ occurs specifically because $Q_M$ is independent of the temperature. In the classical scenario of Landauer's principle, the amount of heat dissipated by the memory increases with the temperature of the environment such that Ineq.~(\ref{landauermemo}) is always fulfilled \cite{maru}. At this point, we call attention to the fact that this protocol {\it does not} represent a violation of Landauer's principle because the initial state of the reservoir is not completely thermal because it encompasses the pure ancilla. In Fig.~\ref{temperature} we display a map of $T_l$ as a function of the position of $\hat{\rho}^{(i)}_{M}$ in the Bloch sphere for states with $r = 0.5$.

In order to make a correspondence between our approach and the classical Landauer scenario, in which one bit of information is erased, we have to consider the state of the memory as a classical bit, i.e., a completely mixed qubit state, $\hat{\rho}^{(i)}_{M} = \hat{\mathbb{I}}_M / 2$. This provides $r_z=r=0$, such that the limit temperature is given by $T_l =  (\mathrm{ln} \, 4)^{-1} \Delta / k_B $. We observe that $T_l$ can be well below room temperature by choosing a small energy level spacing $\Delta$. For example, for Li Rydberg atoms with principal quantum number $n$ near 30, we have for the nearest transitions that $\Delta \approx 10$ cm$^{-1}$ ($\approx 1.986 \times 10^{-22}$ J) \cite{kleppner,saff}, which provides $T_l \approx 10$ K.

\section{Optical simulation} \label{opticalsimulation}

\begin{figure}[h!]
    \centering
    \includegraphics[scale=0.22]{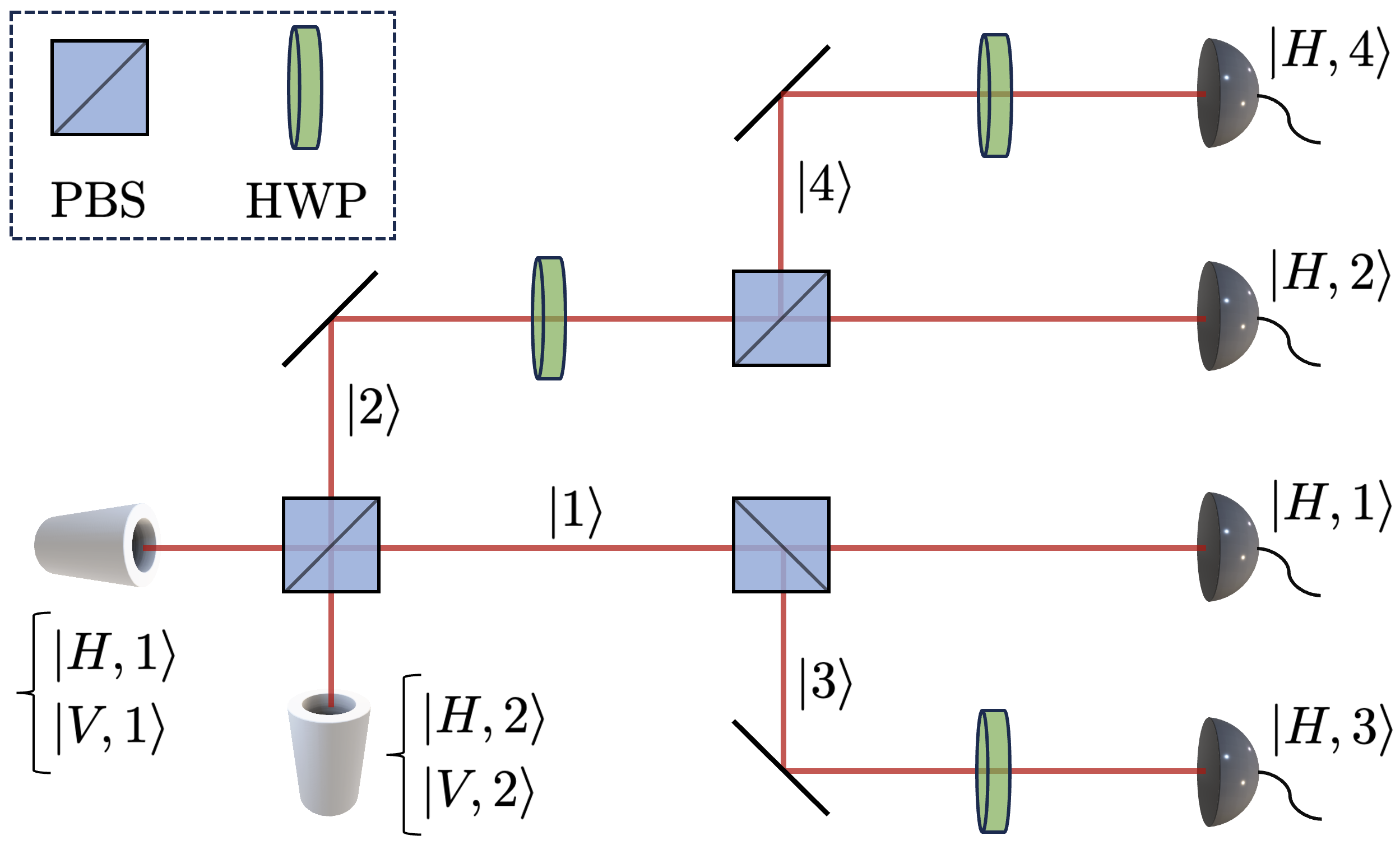}
    \caption{Optical circuit that simulates the erasure operation. The memory state is encoded on the polarization of the photons, which enter the circuit through the paths $\ket{1}$ and $\ket{2}$. The two-qubit state of the reservoir is encoded on these two paths and the paths $\ket{3}$ and $\ket{4}$. With the transformations generated by the arrangement of PBSs and HWPs, it is easy to see that the photons exit the circuit necessarily with horizontal polarization (memory erased), and in a mixture of the four possible paths. The absence of optical losses in the circuit is a reflection of the unitarity of the general process.}
    \label{fig4}
\end{figure}

We now describe a quantum optical simulation of the erasure operation, which is based on the setup sketched in Fig.~\ref{fig4}. Here, the energy levels of the memory are associated with the horizontal and vertical polarizations of the photons that propagate through the optical circuit: $\ket{g_{M}} \leftrightarrow \ket{H}$ and $\ket{e_{M}} \leftrightarrow \ket{V} $. The four states of the reservoir are associated with the four possible paths (optical modes) that the photons can take: $\ket{g_{R}, l_{0}} \leftrightarrow \ket{1}$; $\ket{e_{R}, l_{0}}  \leftrightarrow \ket{2}$; $\ket{g_{R}, l_{1}} \leftrightarrow \ket{3}$  and $\ket{e_{R}, l_{1}} \leftrightarrow \ket{4} $. According to the relations described in Eqs.~(\ref{first}) to~(\ref{fourth}), the circuit is required to make the following transformations: 
\begin{equation}\label{expfirst}
        \ket{H,1} \longrightarrow \ket{H,1},
    \end{equation}
    \begin{equation}\label{expsecond}
        \ket{H,2}\longrightarrow \ket{H,4},
    \end{equation}
\begin{equation}\label{expthird}
        \ket{V,1}\longrightarrow \ket{H,2},
    \end{equation}
    \begin{equation}\label{expfourth}
        \ket{V,2}\longrightarrow \ket{H,3}.
    \end{equation}
These relations show that, independently of the input polarization state (memory) of the photons, and the input path state (reservoir), which is here preselected as a mixture of the states  $\ket{1}$ and $\ket{2}$, the memory ends up necessarily in the pure state $\ket{H}$, which means that it is erased. 

The circuit consists of an arrangement of three polarizing beam splitters (PBSs) and three half-wave plates (HWPs). The PBSs transmit $\ket{H}$ and reflect $\ket{V}$ states, and the HWPs are configured to transform $\ket{H}$ states into $\ket{V}$ states and vice versa \cite{kok}. According to Eq.~(\ref{irstate}), the initial (preselected) path state reads
\begin{equation}
\label{expirstate}
\begin{aligned}
    \hat{\rho}^{(i)}_{\mathrm{path}} = p_{1} \ket{1}\bra{1} + p_{2}\ket{2}\bra{2},
\end{aligned}
\end{equation}
with the statistical weights $p_1$ and $p_2$ simulating a Boltzmann distribution, namely $p_{2} = e^{-\beta \Delta} p_{1}$ and $p_{1} + p_{2} = 1$. Experimentally, the state $\hat{\rho}^{(i)}_{\mathrm{path}}$ is generated by controlling the probabilities of sending photons through the input ports $\ket{1}$ and $\ket{2}$, without phase relation between them.

The complete initial state of the photons is $\hat{\rho}^{(i)} = \hat{\rho}^{(i)}_{p} \otimes \hat{\rho}^{(i)}_{\mathrm{path}}$, with $\hat{\rho}^{(i)}_{p}$ representing the arbitrary polarization state to be erased, i.e.,  
\begin{equation}
\label{opticalimstate}
\hat{\rho}^{(i)}_{p} =  \frac{1}{2} (\hat{\mathbb{I}}_p + r_x \hat{\sigma}_{x} + r_y \hat{\sigma}_{y} + r_z \hat{\sigma}_{z}).
\end{equation}
In this case, the components of the Bloch vector $r_x$, $r_y$ and $r_z$ represent the Stokes parameters of the initial polarization state \cite{james,lukas}, and $\hat{\mathbb{I}}_p$ is the identity operator in the polarization basis $\{ \ket{H}, \ket{V}\}$. 

According to the erasure operation effect on the composite system presented in Sec.~\ref{erasure}, the setup transforms the initial state $\hat{\rho}^{(i)}$ into $\hat{\rho}^{(f)} = \ket{H}\bra{H} \otimes \hat{\rho}^{(f)}_{\mathrm{path}}$, where $\hat{\rho}^{(f)}_{\mathrm{path}}$ is the final path state, which has the same form as that given in Eq.~(\ref{frstate}). Namely, 
\begin{align}
\label{opticalfrstate}
\hat{\rho}^{(f)}_{path} &= \frac{1}{2} \Bigl[ (1 + r_{z})(p_{1}\ket{1}\bra{1} + p_{2}\ket{4}\bra{4}) \nonumber
\\
    &+ (r_{x} - ir_{y})(p_{1}\ket{1}\bra{2} +
p_{2}\ket{4}\bra{3}) \nonumber
\\
    &+ (r_{x} + ir_{y})(p_{1}\ket{2}\bra{1} +
p_{2}\ket{3}\bra{4}) \nonumber
\\
    &+ (1 - r_{z})(p_{1}\ket{2}\bra{2} 
 +p_{2}\ket{3}\bra{3}) \Bigr].
\end{align}

\section{Conclusions and outlook}
\label{conclusions}

We have proposed a quantum channel based on an arrangement of CNOT gates that efficiently erases the information encoded in a qubit memory in contact with
reservoir made up of a thermal qubit and a pure ancilla qubit. As shown in Sec.~\ref{erasure}, the ancilla can be either a non-energetic degree of freedom of the reservoir (e.g. angular momentum) or an independent qubit system. A particular aspect of the approach is that the heat dissipated by the memory depends on its initial state, but not on the temperature of the reservoir thermal qubit [see Eq.~(\ref{hem})]. This fact implies that Landauer's bound is not satisfied [see Ineqs.~(\ref{landauer}) and~(\ref{landauermemo})] when the thermal qubit temperature is above a limit temperature [see Eq.~(\ref{limit})], which can be well below room temperature if we consider the structure of the memory and the thermal qubit based on the nearest transitions of Rydberg atoms \cite{kleppner,saff}. The erasure operation is also illustrated with a proposed linear-optical setup.

Although our erasing scheme can operate in a regime that goes beyond the Landauer limit, we cannot say that it violates Landauer's principle because our reservoir as whole does not start out in a thermal state, which is a requirement of the Landauer scenario. The statement that the erasure of information always produces heat holds true. In fact, as discussed in Subsec.~\ref{thermodynamics} B, the memory always ends up in the ground state by releasing some amount of heat. In this context, the present erasure protocol can also be understood as a perfect cooling process. Our results help clarifying how quantum strategies relate to Landauer's principle, and how they can contribute to the emerging field of green computing \cite{jasch}, which aims to reduce the harmful impact of energy-consuming technologies on the environment.

\section*{Acknowledgments}

\noindent The authors acknowledge support from Coordena{\c c}{\~a}o de Aperfei{\c c}oamento
de Pessoal de N{\'i}vel Superior (CAPES, Finance Code 001) and Conselho Nacional de Desenvolvimento Cient{\'i}fico e Tecnol{\'o}gico (CNPq). BLB acknowledges support from CNPq (Grant No. 307876/2022-5).
\\

\section*{Data availability} 

\noindent All data generated or analysed during this study are included in this published article.

\end{document}